\documentclass{article}

\usepackage[accepted]{aistats2021_modified}
\usepackage{algorithm} % algorithm environment
\usepackage{amssymb, amsmath, amsthm} % math symbols etc.
\usepackage{appendix} % appendix numbering
\usepackage{arydshln} % dashed lines in tables
\usepackage{enumitem} % detailed list control
\usepackage{float} %control over float placement
\usepackage[symbol]{footmisc} % for footnote symbols rather than numbers
\usepackage[colorlinks=true, allcolors=LinkBlue, backref=page,hyperfootnotes=false]{hyperref} % ref links
\usepackage{mathtools} % required by AISTATS
\usepackage{mathrsfs} % mathscr style
\usepackage{multicol} % used for title and author section
\usepackage[round]{natbib} % for citet and citep. AISTATS requires round mode
\usepackage[compact]{titlesec} % control over title spacing
\usepackage[dvipsnames]{xcolor} % for named colors

\setlist{noitemsep, topsep=0pt, leftmargin=!} % for enumitem
\definecolor{LinkBlue}{rgb}{.10, .20, .80} % for hyperref

\newcommand\blfootnote[1]{%
  \begingroup
  \renewcommand\thefootnote{}\footnote{#1}%
  \addtocounter{footnote}{-1}%
  \endgroup
}

\newfloat{algorithm}{tbp}{loa}
\providecommand{\algorithmname}{Algorithm}
\floatname{algorithm}{\protect\algorithmname}

\newtheorem{prop}{Proposition}
\newtheorem{lemma}{Lemma}

\newcommand{\eq}[1]{\begin{align*}#1\end{align*}} % alignable equation*
\newcommand{\Unif}{\operatorname{Unif}}
\newcommand{\Bern}{\operatorname{Bern}}
\newcommand{\N}{\operatorname{N}}
\newcommand{\Expo}{\operatorname{Expo}}
\newcommand{\E}{\operatorname{\mathbb{E}}}

\renewcommand{\P}{\operatorname{\mathbb{P}}}

\newcommand{\R}{\ensuremath{\mathbb{R}}}

\newcommand{\cc}{\:\!} %\, is a thin space, \cc is even thinner

\newcommand{\ty}{\tilde{y}}

\newcommand{\diff}{\, \mathrm{d} } % e.g. \int f(x) \diff x
\newcommand{\TV}{{\sf TV}}

\newcommand{\calX}{\mathcal{X}}
\newcommand{\scrF}{\mathscr{F}}

\newcommand{\BQMI}{\bar{Q}_\mathrm{MI}} % max independent coupling of proposals
\newcommand{\BQMR}{\bar{Q}_\mathrm{MR}} % max reflection coupling of proposals

\newcommand{\BPSQ}{\bar{P}_\mathrm{SQ}} % the (S)tatus (Q)uo MH coupling described in Jacob et al. [2020]
\newcommand{\BPMI}{\bar{P}_\mathrm{MI}} % the (M)aximal full-kernel coupling with (I)ndependent residuals
\newcommand{\BPMR}{\bar{P}_\mathrm{MR}} % the (M)aximal full-kernel coupling with attempted (R)eflection residuals
\newcommand{\BPC}{\bar{P}_\mathrm{C}} % the modified version of BPSQ using (C)onditional accept/reject

\begin{document}

\twocolumn[

\aistatstitle{Maximal Couplings of the Metropolis--Hastings Algorithm} % Full capitalization, per AISTATS
\runningtitle{Maximal Couplings of the MH Algorithm}

\aistatsauthor{John O'Leary\footnotemark[1] \And Guanyang Wang\footnotemark[1] \And Pierre E. Jacob}
\runningauthor{O'Leary,  Wang, \& Jacob}

\aistatsaddress{Harvard University \And Rutgers University \And Harvard University}
\aistatsaddress{  
	\href{mailto:joleary@g.harvard.edu} {joleary@g.harvard.edu}
	\And \href{mailto:guanyang.wang@rutgers.edu}{guanyang.wang@rutgers.edu}
	\And \href{mailto:pjacob@g.harvard.edu}{pjacob@g.harvard.edu}}
\vskip 0.3in
]

\begin{abstract} % AISTATS requires this to be a single paragraph
Couplings play a central role in the analysis of Markov chain Monte Carlo algorithms 
	and appear increasingly often in the algorithms themselves, e.g. in convergence 
	diagnostics, parallelization, and variance reduction techniques. Existing couplings of
	the Metropolis--Hastings algorithm handle the proposal and acceptance steps separately 
	and fall short of the upper bound on one-step meeting probabilities given by the coupling inequality. 
	This paper introduces maximal couplings which achieve this bound while retaining the practical 
	advantages of current methods. We consider the properties of these couplings and examine 
	their behavior on a selection of numerical examples. \blfootnote{\hspace{-1ex}$^*$The first two authors contributed equally to this work.}
\end{abstract}

%%%%%

\vspace{-2ex}
\section{Introduction}
\label{sec:intro}

Markov chain Monte Carlo (MCMC) methods offer a powerful framework for approximating
integrals over a wide range of probability distributions \citep{Brooks2011}.
The Metropolis--Hastings (MH) family of algorithms has proved to be especially
popular, from its original forms \citep{Metropolis1953, Hastings1970} to modern
incarnations such as Hamiltonian Monte Carlo \citep{Duane1987, Neal1993,
	Neal2011}, the Metropolis-adjusted Langevin algorithm \citep{Roberts1996a}, and
particle MCMC \citep{andrieu2010particle}. In many settings MH methods present
an attractive mix of effectiveness, flexibility, and transparency.

Couplings of MCMC transition kernels have long played a role in the analysis
and diagnosis of their convergence 
\citep{Rosenthal1995, Johnson1996, Johnson1998, jerrum1998mathematical, Rosenthal2002, Biswas2019},
as a way of obtaining perfect samples or unbiased estimators 
\citep{Propp1996,Neal2017,Glynn2014,Heng2019,Jacob2020,Middleton2019, Middleton2020},
and as a variance reduction technique
\citep{neal2001improving,Goodman2009,piponi2020hamiltonian}.
Such couplings are usually required to make the chains meet in finite time, with smaller meeting times associated with tighter bounds and greater precision or computational efficiency. In
practice it is also essential for couplings to be implementable, in the sense
that they require no extra knowledge about the target distribution beyond the requirements of the
underlying MCMC algorithm.

We take up the question of coupling continuous state-space MH chains, following
\citet{Johnson1996,Johnson1998} and \citet{Jacob2020}. In Section \ref{sec:cpl} we define
our setting and review existing methods. In Section \ref{sec:cpl:fullkern} we introduce a set of implementable
couplings which achieve the largest possible meeting probability at each
iteration. These are the first known MH couplings with this property. We
compare these algorithms with existing methods and introduce refinements that
combine maximality with the benefits of the status quo. In Section~\ref{sec:num}
we apply our couplings to two numerical examples, gaining further insight into
their properties and behavior. Finally, in Section \ref{sec:discussion} we consider open questions
and next steps.

%%%%%

\section{Metropolis--Hastings Couplings} % Full capitalization, per AISTATS
\label{sec:cpl}

%%%

\subsection{Setting and Definitions}
\label{sec:cpl:defns}

We write ${x \wedge y = \min(x,y)}$, $x \vee y = \max(x,y)$, $\Unif$ for the
uniform distribution on $[0,1]$, and $\Bern(\alpha)$ for the Bernoulli distribution on $\{0,1\}$ with ${\P(\Bern(\alpha)=1)=\alpha}$.

\begin{figure}
	\includegraphics[width=\linewidth]{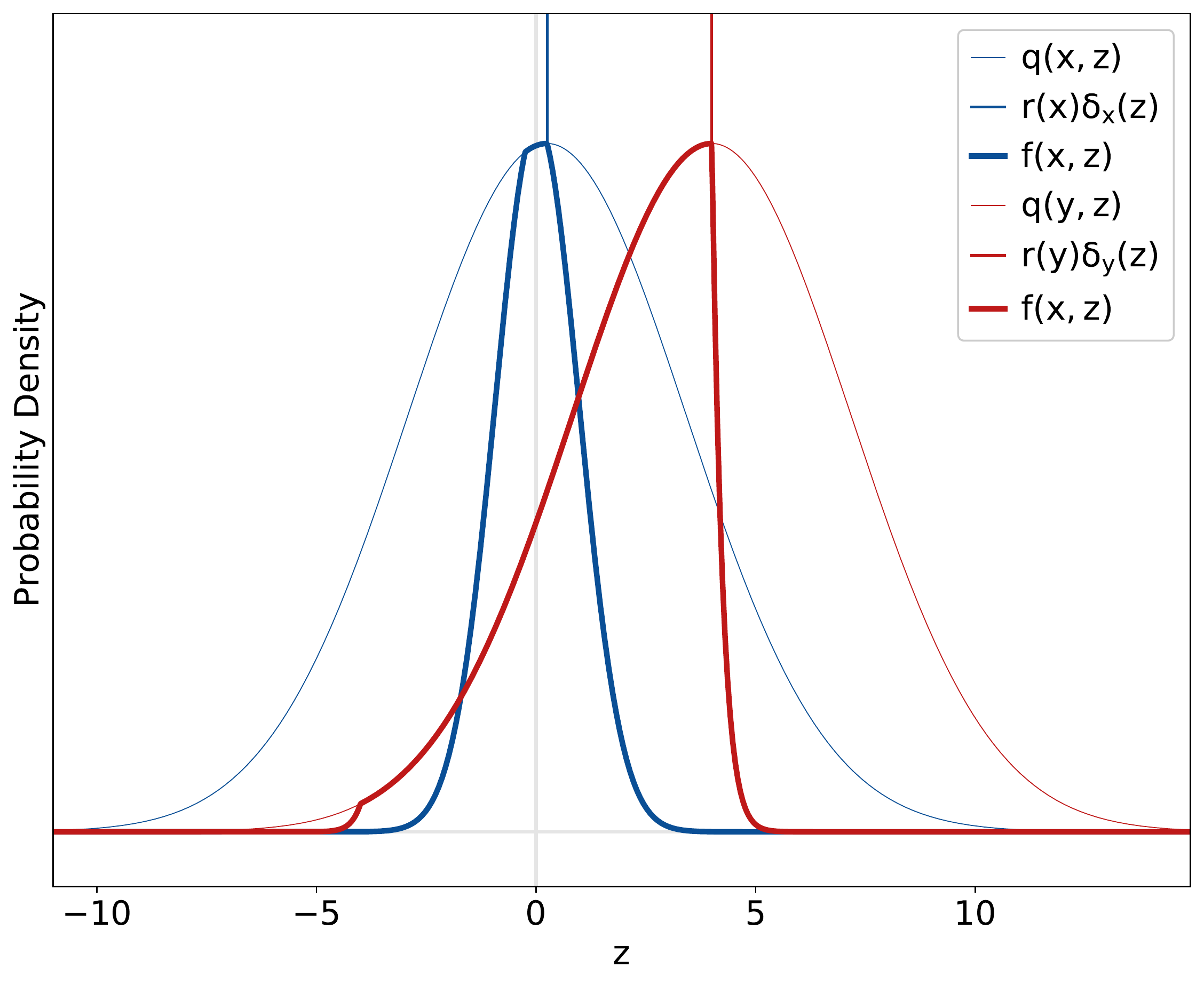}
	\caption{Proposal densities $q$ and MH transition densities $f$, with 
		${\pi = \N(0,1)}$, $x=1/4$, $y=4$, and $Q(z,\cdot) = \N(z, 10)$.
		$P(x,\cdot)$ and $P(y,\cdot)$ also contain point masses
		with weights ${r(x) \approx 0.69}$ and ${r(y) \approx 0.47}$, respectively.
		The algorithms described in this paper construct couplings of such transition kernels, 
		sometimes by way of proposal distribution couplings.
		\label{fig:qandf}}
\end{figure}

Let $P$ be a Markov transition kernel with stationary distribution $\pi$ on $(\calX, \scrF)$, a Polish space equipped with the standard Borel $\sigma$-algebra.   For ${x \in \calX}$ and ${A \in \scrF}$,
$P(x,A)$ denotes the probability of a transition from $x$ to $A$. We focus on
MH-like kernels $P$, characterized by the property that we can obtain ${X \sim
	P(x,\cdot)}$ by drawing a proposal ${x' \sim Q(x,\cdot)}$ and an acceptance indicator ${B \sim \Bern(a(x,x'))}$ and
setting ${X = B x' + (1-B) x}$. We assume that for all $x \in \calX$, $Q(x,\cdot)$ has density $q(x,\cdot)$ with respect to a base measure on $(\calX, \scrF)$.
We also assume the proposal distribution is non-atomic, so that $Q(x,\{y\}) = 0$ for all $x,y \in \calX$.
The acceptance rate under MH will be ${a(x,x') = 1 \wedge
	\tfrac{\pi(x') q(x',x)}{\pi(x) q(x,x')}}$, and we allow for alternatives such as Barker's algorithm \citep{Barker1965}.
For $x' \neq x$ we define ${f(x, x') := q(x, x') a(x, x')}$ and ${r(x) := 1 - \int f(x, x') \diff
	x'}$, so that $P(x,\cdot)$ has density $f(x,x')$ except for an atom where $P(x,\{x\})
= r(x)$. See Figure \ref{fig:qandf} for an illustration of a pair of proposal and transition distributions.

A probability distribution $\gamma$
on $\calX \times \calX$ is a coupling of distributions $\mu$ and $\nu$ on
$(\calX, \scrF)$ if ${\gamma(A \times \calX) = \mu(A)}$ and ${\gamma(\calX \times A) = \nu(A)}$
for any ${A \in \scrF}$. Let $\Gamma(\mu, \nu)$ be the set of all
couplings of $\mu$ and $\nu$.
We say that a joint kernel $\bar P$ is a coupling of $P$ with itself and write $\bar P \in \Gamma(P,P)$
if $\bar P((x,y,), \cdot) \in \Gamma( P(x,\cdot), P(y, \cdot))$ for any $x, y \in \calX$.
Similar definitions apply to couplings $\bar Q$ of proposal distributions and couplings $\bar B$ of acceptance indicators.

The coupling inequality \citep[Proposition 4.7]{Levin2017} states that 
the meeting probability ${\P_{(X,Y)
		\sim \gamma}(X=Y) \leq 1 - \lVert \mu - \nu \rVert_\TV}$
for any coupling $\gamma \in \Gamma(\mu, \nu)$.
Here ${\lVert \mu -
	\nu\rVert_\TV = \sup_{A \in \scrF} | \mu(A) - \nu(A) |}$ is the total variation
distance.  A coupling that achieves this bound is said to be maximal, and we
write $\Gamma^\mathrm{max}(\mu,\nu) \subset \Gamma(\mu,\nu)$ for the set of
maximal couplings of $\mu$ and $\nu$.

%%%

\subsection{Status Quo: the Heuristic Coupling $\BPSQ$}
\label{sec:cpl:heuristic}

\begin{figure}[t]
	\includegraphics[width=\linewidth]{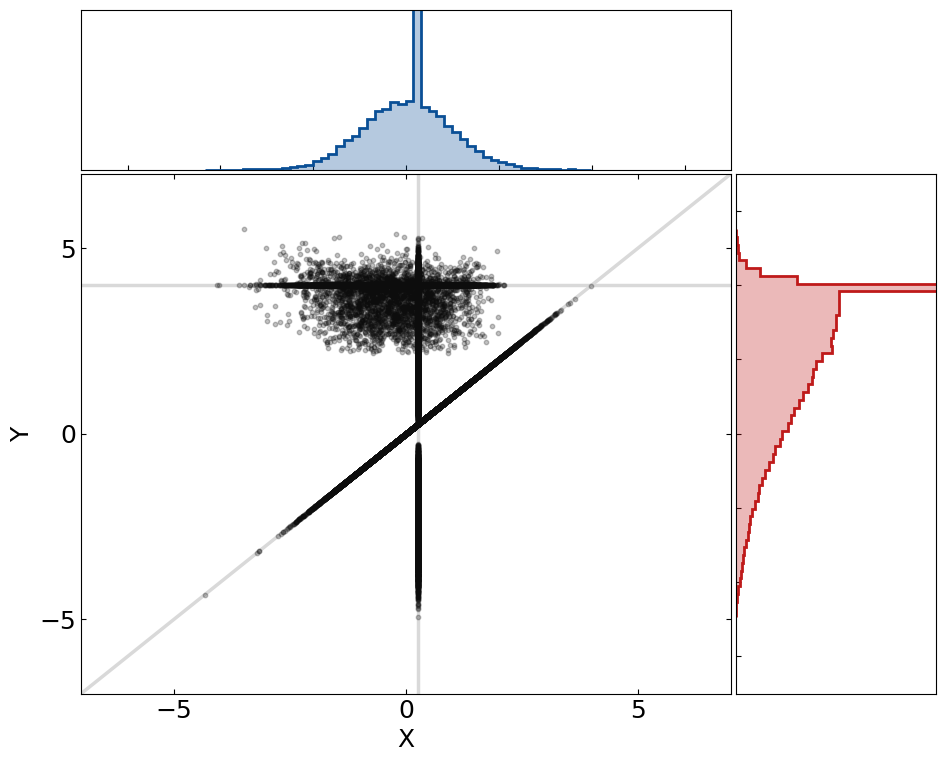} %pdf also available but it slows down rendering
	\caption{Draws from the coupling $\BPSQ((x,y),\cdot)$ using $\BQMI$ and the same parameters as 
		in Figure \ref{fig:qandf}. The grey lines indicate $X=x$, $Y=y$, and $X=Y$, while the histograms show the marginal distributions of $X$ and~$Y$.
		\label{fig:qscatter}}
\end{figure}
We begin by describing the state-of-the-art coupling of MH transition kernels, first introduced in \citet{Johnson1998}.
Recall that draws from a MH-like kernel $P(x,\cdot)$ can be obtained via a proposal ${x' \sim Q(x,\cdot)}$ which is accepted for $X$ with probability $a(x,x')$. Thus a simple way to couple $P$ with itself is to draw coupled proposals ${(x',y') \sim \bar{Q}((x,y),\cdot)}$ and accept or reject these according to coupled indicators ${(B_x, B_y) \sim \bar{B}((x,y),(x',y'))}$, where
${\bar Q \in \Gamma(Q,Q)}$ and $\bar B((x,y),(x',y')) \in \Gamma \big(\! \Bern(a(x,x')), \Bern(a(y,y')) \big)$.
We refer to this coupling as $\BPSQ$ and summarize it in Algorithm \ref{alg:bpsq}.

\begin{algorithm}[b]
	\caption{Draw $(X,Y) \sim \BPSQ((x,y),\cdot)$ \label{alg:bpsq}}
	\begin{enumerate}
		\item Draw $(x',y') \sim \bar Q((x,y),\cdot)$
		\item Draw $(B_x, B_y) \sim \bar B((x,y),(x',y'))$
		\item Set $X \cc = \cc B_x\, x' + (1-B_x) x$ and $Y \cc = \cc B_y\, y' + (1-B_y) y$
		\item Return $(X,Y)$
	\end{enumerate}
\end{algorithm}

For the chains to meet in finite time we need ${ \P(X=Y \mid x,y) > 0 }$ from at least some states $(x,y)$, 
which in turn requires ${\P(x'=y' \mid x,y) >0}$.
To obtain this we take the joint proposal distribution $\bar Q$ to be a maximal coupling
of $Q$ with itself. We can draw from such a coupling by sampling $x' \sim Q(x,\cdot)$,
using this as a rejection sampling proposal for $Q(y, \cdot)$, and drawing $y'$
in a specified way if this $x' = y'$ proposal is rejected. 
This approach achieves the coupling inequality upper bound ${\P(x'=y' \mid x,y) = 1 - \int q(x, z) \wedge q(y,z) \diff z}$.

\begin{algorithm}
	\caption{Draw $(x',y') \sim \BQMI((x,y),\cdot)$ \label{alg:bqmi}}
	\begin{enumerate}
		\item Draw $x' \sim Q(x,\cdot)$ and $U \sim \Unif$
		\item If $U q(x,x') \leq q(y,x')$, set $y' = x'$
		\item Else
		\begin{enumerate}
			\item Draw $\ty \sim Q(y, \cdot)$ and $V \sim \Unif$
			\item If $V q(y, \ty) > q(x, \ty)$, set $y' = \ty$
			\item Else go to 3(a)
		\end{enumerate}
		\item Return $(x',y')$
	\end{enumerate}
\end{algorithm}

A simple example of this method is the maximal coupling with independent residuals $\BQMI$,
introduced in \citet{Vaserstein1969} and called the $\gamma$-coupling in \hbox{\citet[chap.~1.5]{Lindvall1992}}.
It has the property that $x'$ and $y'$ are independent when $x' \neq
y'$. Algorithm \ref{alg:bqmi} describes how to draw from this coupling,
and Figure \ref{fig:qscatter} illustrates
a set of draws $(X,Y) \sim \BPSQ((x,y),\cdot)$ based on proposals from $\BQMI$.

When $\calX = \R^d$ and $Q$ is spherically symmetric, the maximal
coupling with reflection residuals, $\BQMR$,
is often a better alternative.
This coupling was introduced in the context of Hamiltonian and Langevin methods \citep{Bou-Rabee2018, eberle2019couplings} and has its origins in the analysis of continuous-time processes \citep{Lindvall1986, Eberle2011}.
$\BQMR$ is identical to $\BQMI$ for $x' = y'$.
When $x' \neq y'$ the coupling $\BQMR$ sets $y' =
T_{xy}(x')$, where $T_{xy}(x') = y + (I - 2 e e')(x'-x)$ 
and $e = (y-x)/\lVert y-x\rVert$. We define $T_{yx}(y')$ similarly. Note that the transformation $z \to (I -
2 ee')z$ reflects the $e$ component of $z$ while leaving the $e^\perp$ component
fixed.

A standard choice for the acceptance indicators $(B_x, B_y) \sim \bar B((x,y),(x', y'))$ 
is the unique maximal coupling of $\Bern(a(x,x'))$ and $\Bern(a(y,y'))$, which can be 
realized by drawing $U \sim \Unif$ and setting ${B_x = 1(U \leq a(x,x'))}$ and 
${B_y = 1(U \leq a(y,y'))}$. Among couplings of these distributions, this one yields the maximal probability
${\P(B_x= B_y=1 \mid x', y')} = {a(x,x') \wedge a(y,y')}$.
Other objectives, such as the minimization of $\E[ \lVert X-Y \rVert \mid x',y' ]$, are also possible.

\begin{figure}
	\includegraphics[width=\linewidth]{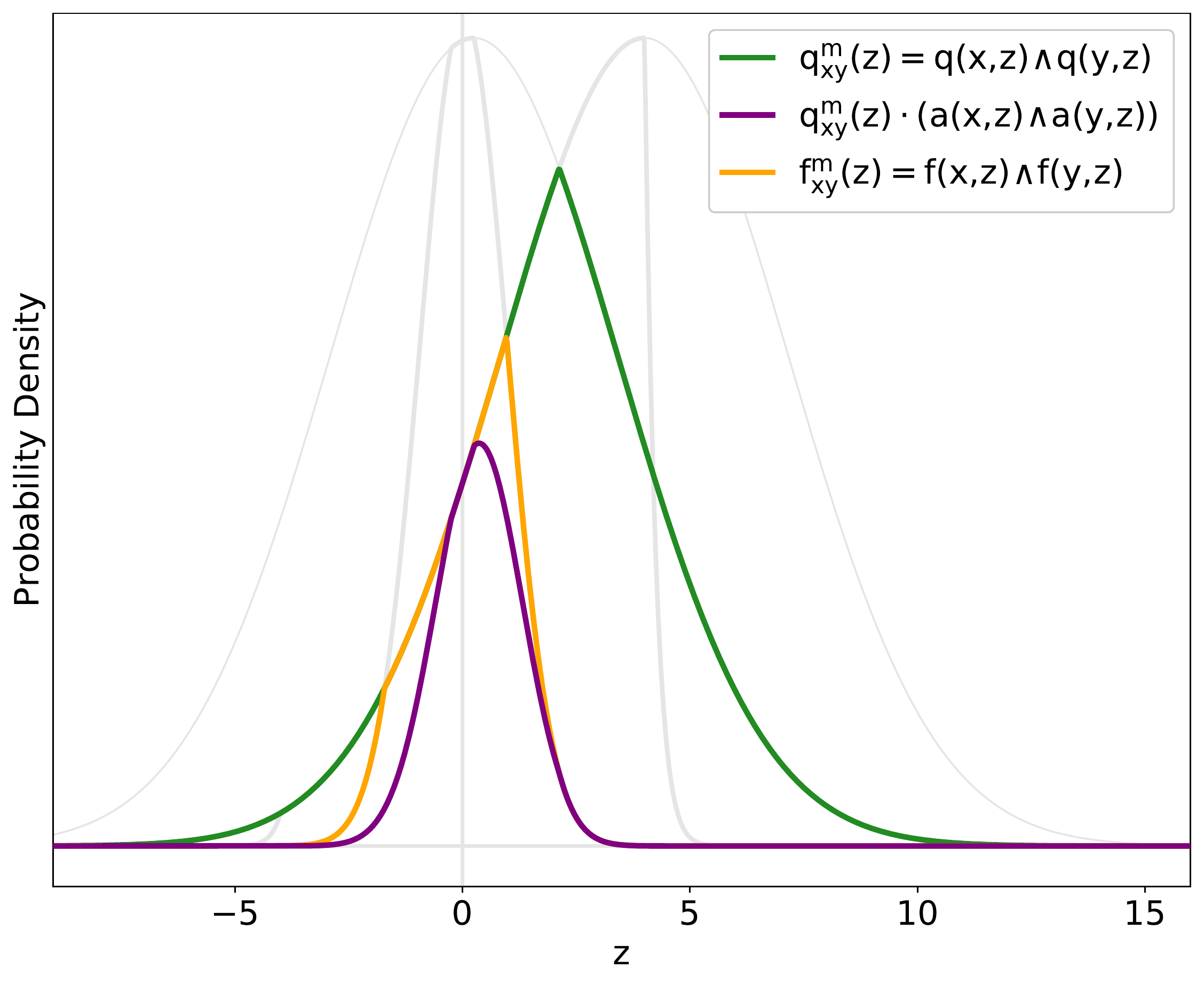}
	\caption{Meeting densities based on the same parameters as in Figure \ref{fig:qandf}.
		The green line shows the density of ${x' = y' = z}$ under a maximal coupling $\bar Q$. The purple line
		shows the density of $X=Y=z$ under $\BPSQ$ assuming the use of a maximal coupling $\bar B$.
		Finally, the orange line shows the density of $X=Y=z$ under a maximal coupling $\bar P$. The
		inequalities suggested here hold in general.
		\label{fig:maxprob} }
\end{figure}

The coupling $\BPSQ$ obtained when $\bar Q$ is a maximal coupling of proposal distributions 
and $\bar B$ is the maximal coupling of acceptance indicators 
can yield a relatively high chance of ${X=Y}$ given the current state pair $(x,y)$,
but it typically falls short of the theoretical upper bound.
Under $\BPSQ$, the probability of $X=Y$ is ${ \int\! \big(q(x,z) \wedge q(y,z)
	\big) \big( a(x,z) \wedge a(y,z) \big) \diff z}$. However Lemma
\ref{lem:kernel_tv} implies that the coupling inequality bound is
$\int\! \big(q(x,z) a(x,z)\big) \wedge \big(q(y,z) a(y,z)\big) \diff z$,
which is always an equal or larger quantity.
Figure \ref{fig:maxprob} illustrates the gap
between the meeting probabilities under $\BPSQ$ and under
any maximal coupling $\bar P$.
Note that these probabilities will coincide when either $q(x,z)$ or $a(x,z)$ 
does not depend on $x$, e.g. for the independence sampler.

\begin{lemma}
	\label{lem:kernel_tv}
	Let $P$ be an MH-like transition kernel as defined above. Then for $x\neq y$,
	${\lVert P(x,\cdot) - P(y, \cdot)\rVert_\TV} = {1 - \int f(x,z) \wedge f(y,z) \diff z}$.
\end{lemma}

\vspace{-1em}
\begin{proof}
	Let $C_{xy} = \{z: f(x,z) > f(y,z)\} \cup \{x\} \setminus \{y\}$ and similarly for $C_{yx}$. 
	We have ${ \lVert P(x,\cdot) - P(y,\cdot)\rVert_\TV} = { \sup_A| P(x,A)-P(y,A)| }$,
	and ${ | P(x,A) - P(y,A) | } = { |\int_A f(x,z)\! - \!  f(y,z) \diff z + 1(x \in A)r(x)\! - \! 1(y \in A) r(y)|. }$
	We must have either $A = C_{xy}$ or $A=C_{yx}$ in the supremum
	above. Both yield the same value, and so 
	$\lVert P(x,\cdot) - P(y, \cdot)\rVert_\TV = 1 - \int f(x,z) \wedge f(y,z) \diff z$.
\end{proof}

We want to maximize the coupling probability at each step and reduce 
typical meeting times, so the factors above lead us to ask if
there are couplings $\bar P$ which are both implementable and maximal.
Our contribution answers this question in the affirmative.

%%%

\section{Maximal Couplings}
\label{sec:cpl:fullkern}

We now introduce a collection of implementable couplings ${\bar P \in
	\Gamma^\mathrm{max}(P,P)}$ starting from an arbitrary MH-like transition
kernel $P$. 
We consider two approaches.
In Sections \ref{sec:cpl:bpmi} and \ref{sec:cpl:bpml} we show how to couple Markov transition kernels
without explicitly coupling their underlying proposal or acceptance distributions. 
We refer to these  as full-kernel couplings.  
In Section \ref{sec:cpl:conditional}, we
modify Algorithm \ref{alg:bpsq} to maximize the probability of accepting
proposals $x'=y'$ at the expense of decreasing the acceptance probabilities
when $x' \neq y'$. 
%This approach yields a maximal coupling $\bar P$ from any
%maximal coupling $\bar Q$ of proposal distributions.

\begin{algorithm}[b]
	\caption{Draw $(X,Y) \sim \BPMI((x,y),\cdot)$ \label{alg:bpmi}}
	\begin{enumerate}
		\item Draw $X \sim P(x, \cdot)$ and $U \sim \Unif$
		\item If $X \neq x$ and $U \, f(x, X) \leq f(y, X)$, set $Y = X$
		\item Else
		\begin{enumerate}
			\item Draw $\ty \sim P(y, \cdot)$ and $V \sim \Unif$
			\item If $\ty = y$, set $Y=\ty$
			\item If $\ty \neq y$ and $V \, f(y, \ty) > f(x, \ty)$, set $Y = \ty$
			\item Else go to 3(a)
		\end{enumerate}
		\item Return $(X,Y)$
	\end{enumerate}
\end{algorithm}

\subsection{$\BPMI$, a Full-kernel Coupling with Independent Residuals}
\label{sec:cpl:bpmi}

Our first full-kernel coupling, which we will refer to as $\BPMI$, is inspired by
the procedure described in Algorithm \ref{alg:bqmi} for drawing from the
coupling $\BQMI$ of proposal distributions. A key difference between coupling proposal distributions and transition kernels is that by assumption the proposal distributions are non-atomic and absolutely continuous with respect to an underlying measure, 
while MH-like kernels $P$ can have a point mass at the current state. Therefore our rejection sampling
procedure must be modified to yield draws $X$ and $Y$ with the correct marginal
distributions.

\begin{figure}[t]
	\includegraphics[width=\linewidth]{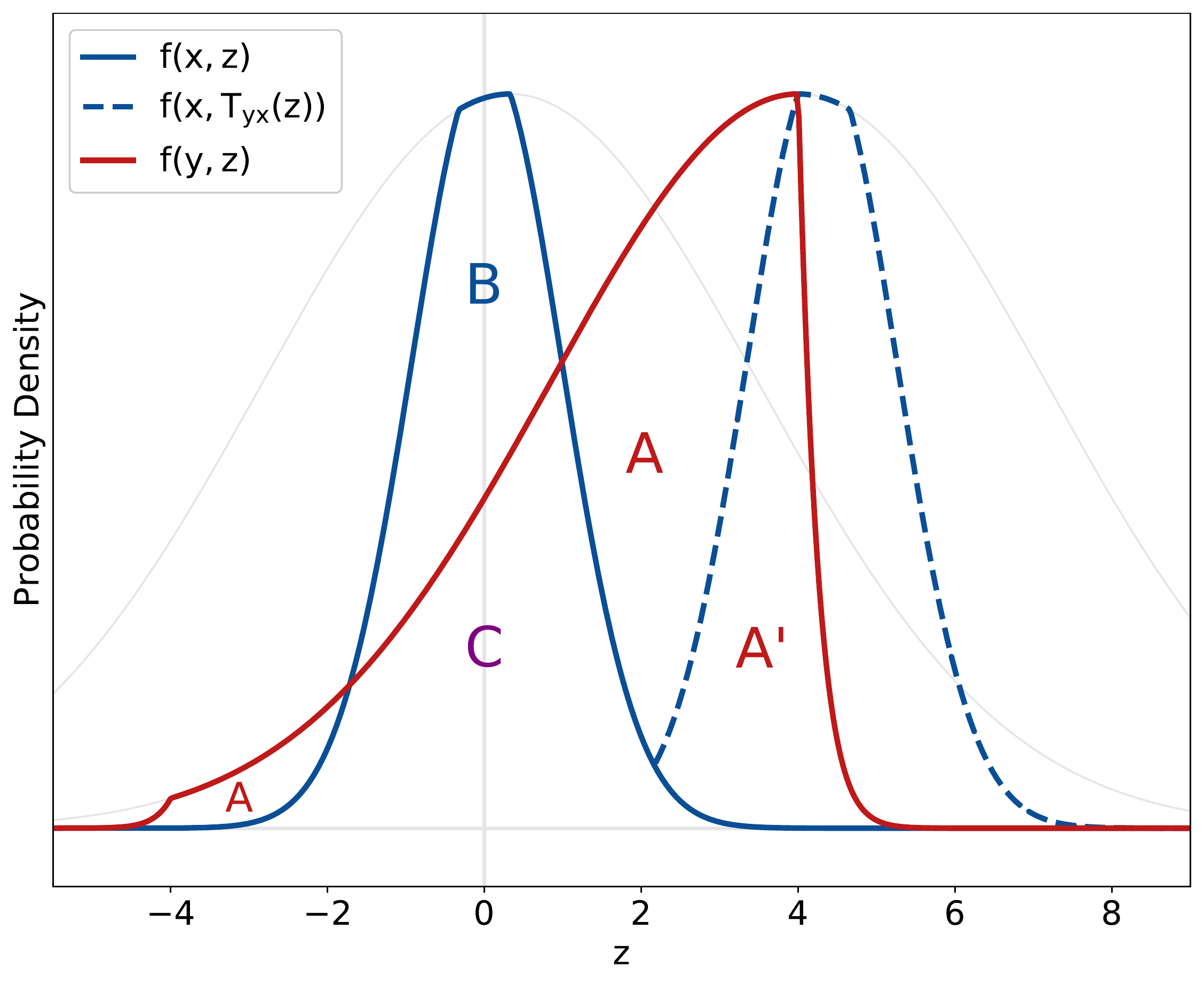}
	\caption{Rejection sampling regions for Algorithms \ref{alg:bpmi} and \ref{alg:bpmr}
		based on the same parameters as Figure \ref{fig:qandf}. For both options, meeting occurs on samples
		drawn from region $C$. \label{fig:coupling_area}} 
\end{figure}

Algorithm \ref{alg:bpmi} contains this modification, which we illustrate in Figure \ref{fig:coupling_area}. 
The blue and red curves give the continuous parts of $P(x,\cdot)$ and
$P(y,\cdot)$, respectively. We think of the algorithm as sampling uniformly
from the region under the graph along with point masses at $x$ and $y$. If the $X$
draw falls in $C$ we can use the same point for both chains, and otherwise we use
rejection sampling to obtain a $Y$ draw from $A \cup A' \cup \{y\}$. Meetings
occur exactly on draws from $C$, and thus it occurs with probability $\int f(x,z) \wedge
f(y,z) \diff z$. See Appendix~\ref{sec:app:bpmiproof} for a detailed proof that
$\BPMI((x,y),\cdot)$ is a maximal coupling of $P(x,\cdot)$ and $P(y,\cdot)$ and
an analysis of the computation cost of this algorithm.

%%%

\subsection{$\BPMR$, a Full-kernel Coupling with Reflection Residuals}\label{sec:cpl:bpml}

All maximal couplings of a given $P$ have equal meeting probability, but some
perform better than others. Although $\P(X=Y\mid x,y)$ is maximal under $\BPMI((x,y),\cdot)$,
$X$ and $Y$ are independent when ${X \neq
	Y}$. This creates a strong tendency for $\lVert Y-X\rVert$ to grow with the
dimension of the state space, as seen in the experiments of
\citet{Jacob2020}. Even under a maximal coupling, $\P(X=Y \mid x,y)$ can be
small until $x$ and $y$ are close. Thus the time for a pair of coupled chains
to meet depends on both the meeting probability from each state pair and the
degree of contraction between chains when meeting does not occur.

The advantages of using $\BQMR$ vs. $\BQMI$ in Algorithm
\ref{alg:bpsq} appear to be due to this consideration. See Section \ref{sec:examples:mvn} below for numerical justifications.
It also motivates the following full-kernel maximal coupling with reflection
residuals, which we refer to as $\BPMR$ and describe in detail in Algorithm
\ref{alg:bpmr}. For this algorithm define ${f^m_{xy}(z):= f(x,z) \wedge
	f(y,z)}$, ${\tilde{f}^r_{xy}(x') := f(x,x') - f_{xy}^m(x')}$, and likewise for
$\tilde{f}^r_{yx}(y')$. Finally set ${\tilde{f}^t_{yx}(y') :=
	\tilde{f}^r_{yx}(y') - \tilde{f}^r_{yx}(y') \wedge \tilde{f}^r_{xy}(T_{yx}(y'))}$, 
the $y$ residual after reflection evaluated at $y'$.

% The following might benefit from some clarification
Figure \ref{fig:coupling_area} provides some intuition into the behavior of
$\BPMR$. The first step of $\BPMR$ is identical to $\BPMI$ in that it attempts
to draw from the region $C$, and a meeting occurs if this is successful.
Otherwise $\BPMR$ proposes the reflected point $T_{xy}(X)$ for $Y$, which
succeeds if this point falls in the region $A'$. If all else fails, we use
rejection sampling to obtain a $Y$ draw from $A \cup \{y\}$. See 
Appendix~\ref{sec:app:bpmrproof} for a detailed proof of the validity and
maximality, and an analysis of the computation cost of $\BPMR$.

\begin{algorithm}[t]
	\caption{Draw $(X,Y) \sim \BPMR((x,y),\cdot)$ \label{alg:bpmr}}
	\begin{enumerate}[itemindent=4pt]
		\item Draw $X \sim P(x, \cdot)$ and $U \sim \Unif$
		\item If $X \neq x$ and $U f(x, X) \leq f(y, X)$, set $Y = X$
		\item Else
		\begin{enumerate}
			\item Set $\ty = T_{xy}(X)$ and draw $V \sim \Unif$
			\item If $X \cc \neq \cc x$ and $V f^r_{xy}(X) \cc \leq \cc \tilde f^r_{yx}(\ty) $, set $Y\cc = \cc \ty$
			\item Else
			\begin{enumerate}
				\item Draw $\ty \sim P(y,\cdot)$ and $W \sim \Unif$
				\item If $\ty = y$, set $Y=\ty$
				\item If $\ty \cc \neq \cc y$ and $W f(y,y') \cc \leq \cc \tilde f^t_{yx}(\ty)$, set $Y \cc = \cc \ty$
				\item Else go to 3(c)i.
			\end{enumerate}
		\end{enumerate}
		\item Return $(X,Y)$
	\end{enumerate}
\end{algorithm}

%%%

\subsection{Maximal Coupled Transitions from Maximally Coupled Proposals with $\BPC$}\label{sec:cpl:conditional}

While the coupling $\BPMR$ sometimes outperforms $\BPMI$, 
it also has a few limitations. First, the reflection proposal
strategy requires a high degree of reflection symmetry between the
distributions of $X$ and $Y$ conditional on $X\neq Y$ to be successful.
For example in the setting of Figure \ref{fig:coupling_area}, the reflection proposal
has more than a 50\% chance of being rejected.
Without such a symmetry $X$ and $Y$ will tend to be conditionally independent,
resulting in poor contraction between chains when meeting does not occur.

Second, any full-kernel coupling is constrained to work directly with the complicated
and irregular geometry of $P$ rather than the simple and often tractable
form of the proposal distribution $Q$. In contrast to the range of couplings and 
optimal transport strategies available
when a standard distribution is used for $Q$, it appears
to be more difficult to design high-performance couplings directly in terms of
the associated transition kernels $P$.

\begin{algorithm}[b]
	\caption{Draw $(X,Y) \sim \BPC((x,y),\cdot)$ \label{alg:bpc}}
	\begin{enumerate}
		\item Draw $(x', y') \sim \bar Q((x,y),\cdot)$
		\item If $x' = y'$ set $\bar B = \bar B_1$, else set $\bar B = \bar B_2$
		\item Draw $(B_x, B_y) \sim \bar B((x,y),(x',y'))$
		\item Set $X \cc = \cc B_x\, x' + (1-B_x) x$ and $Y \cc = \cc B_y\, y' + (1-B_y) y$
		\item Return $(X,Y)$		
	\end{enumerate}
\end{algorithm}

\begin{figure}
	\includegraphics[width=\linewidth]{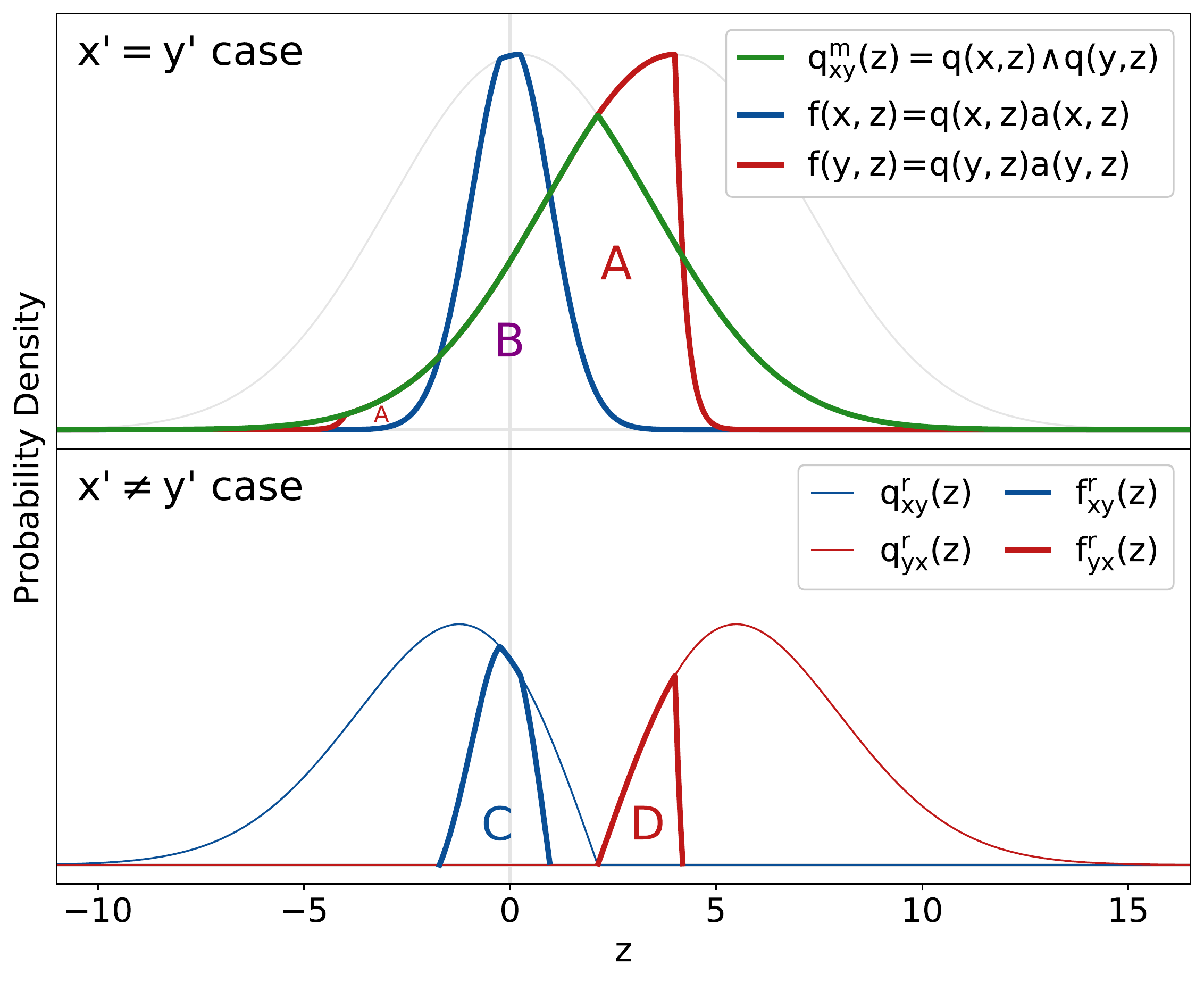}
	\caption{Distributions used in Algorithm \ref{alg:bpc} based on the same parameters as Figure \ref{fig:qandf}. Upper pane: draws ${x'=y'}$ follow $q^m$ and are used as proposals for $f(x,\cdot)$ and $f(y,\cdot)$. Lower pane: Draws $x' \neq y'$ follow the proposal residuals $q^r$ and are used as rejection sampling proposals for the transition kernel residuals $f^r$.
		\label{fig:bpc}}
\end{figure}

This motivates the next coupling, which allows the use of any coupling of $Q$
while still possibly achieving the maximal coupling probability identified in Lemma \ref{lem:kernel_tv}.
% combines the strengths of each of the options described so far.
We refer to this new algorithm as $\BPC$, since it resembles $\BPSQ$ up to the
conditional use of one or another Bernoulli coupling $\bar B$ depending on
whether or not a meeting is proposed. Together with the definitions below, Algorithm \ref{alg:bpc}  shows how to draw from $\BPC$. We can choose a maximal $\bar Q$ such as $\BQMI$ or $\BQMR$ 
to obtain a maximal $\bar P$, although we also obtain a valid $\bar{P} \in \Gamma(P,P)$ for non-maximal choices of $\bar Q$.

To define $\BPC$, let $q^m_{xy}(\cdot)$ be the density of $\bar Q((x,y), \cdot)$
on $\Delta = \{(z,z) : z \in \cal X\}$. We establish the existence and properties of $q^m_{xy}$ in
Lemma \ref{lem:qm}, below. Define the proposal residual 
${q^r_{xy}(x') := {q(x,x') - q_{xy}^m(x')}}$
and the transition kernel residual 
${f_{xy}^r(x') :=0 \vee (f(x,x') - q_{xy}^m(x'))}$,
and similarly for $q^r_{yx}(y')$ and $f^r_{yx}(y')$.
We illustrate these functions in Figure \ref{fig:bpc}.

\begin{lemma}
	\label{lem:qm} If $\bar Q \in \Gamma(Q,Q)$ then there exists a density $q^m_{xy}$ for $\bar Q((x,y), \cdot)$ on $\Delta$. If $\bar{Q}$ is a maximal coupling, then $q^m_{xy}(z) = q(x,z) \wedge q(y,z)$ for almost all $z$.
\end{lemma}

\vspace{-1em}
\begin{proof}
	Let $\lambda$ denote the base measure on $(\calX, \scrF)$
	and let $\lambda_\Delta$ be its push-forward to $\Delta$ by the map ${z \mapsto (z,z)}$.
	Since $\calX$ is a Polish space, we have
	${A_\Delta := \{(z,z) : z \in A\} \in \scrF \otimes \scrF}$
	for any ${A \in \scrF}$.
	Thus $\bar Q((x,y), \cdot)$ induces a sub-probability $\bar Q_\Delta$ on~$\Delta$.
	Also ${ \bar Q_\Delta \ll \lambda_\Delta }$, 
	since if ${\lambda_\Delta(A_\Delta) = \lambda(A) = 0}$ then
	${\bar Q_\Delta(A_\Delta) \leq \bar Q((x,y), A \times \calX) = Q(x, A) = 0}$.
	The Radon–-Nikodym theorem then guarantees the existence of 
	an integrable function ${ q^m_{xy}: \calX \to [0,\infty) } $ such that ${\bar Q((x,y), A_\Delta) = \int_A q^m_{xy}(z) \diff z}$.
	
	Next, we claim ${q^m_{xy}(z) \leq q(x,z) \wedge q(y,z)}$ for $\lambda$-almost all $z$.
	Let ${ A := \{z : q^m_{xy}(z) > q(x,z) \wedge q(y,z) \} \in \scrF }$,
	${A_x := \{z : q^m_{xy}(z) > q(x,z) \} \in \scrF}$, and likewise for $A_y$.
	Since $A = A_x \cup A_y$, $\lambda(A) > 0$ implies $\lambda(A_x) > 0$ or $\lambda(A_y) > 0$.
	If $\lambda(A_x)>0$ then ${\bar Q((x,y), A_x \times \calX)} = {Q(x, A_x)} < {\bar Q((x,y), (A_x)_\Delta)}
	\leq {\bar Q((x,y), A_x \times \calX)}$, a contradiction.  
	The case $\lambda(A_y) > 0$ similarly implies a contradiction, 
	so we conclude that $\lambda(A)=0$.
	
	Finally, if $\bar Q$ is maximal, a total variation computation similar to that of Lemma \ref{lem:kernel_tv}
	shows that $\bar Q((x,y), \Delta) = \int q(x,z) \wedge q(y,z) \diff z$.
	Combining this with the above implies $q^m_{xy}(z) = q(x,z) \wedge q(y,z)$ for $\lambda$-almost all $z$.
\end{proof}

Resuming our definition of $\BPC$, we set acceptance probabilities
$b_{xy}(x') := {1\wedge(f(x,x')/q_{xy}^m(x'))}$ if ${q_{xy}^m(x') > 0}$ or else
${b_{xy}(x') :=1}$, ${c_{xy}(x') := f^r_{xy}(x')/q^r_{xy}(x')}$ if
$q^r_{xy}(x') =0$ or else $c_{xy}(x') := 1$, and likewise for $b_{yx}(y')$ and
$c_{yx}(y')$. When $x'=y'$ we require that the acceptance
indicator pair $(B_x, B_y)$ follows the maximal coupling $\bar B_1$ of
$\Bern(b_{xy}(x'))$ and $\Bern(b_{yx}(y'))$. When ${x' \neq y'}$ we require that
$(B_x, B_y)$ follows any coupling
${\bar B_2 \in \Gamma( \Bern(c_{xy}(x')), \Bern(c_{yx}(y')))}$. Simulation
results suggest that maximal couplings for $\bar B_2$ perform well,
but optimal transport couplings which aim to minimize $\lVert X-Y\rVert$ are also attractive
in this setting.

When $\bar Q$ proposes a meeting $x' = y'$, we think of $\BPC$ as using these
values as rejection sampling proposals for the transition distributions $f(x,z)$ and $f(y,z)$.
This results in a higher marginal acceptance rate than we would have under MH.
On the other hand when $x' \neq y'$. we think of this method as falling back to rejection 
sampling of the residual distributions $f$ from the residuals distributions of $q$, both after 
the removal of $q_{xy}^m$. This produces a lower marginal acceptance rate than MH, 
exactly counterbalancing the above.
In Proposition~\ref{prop:bpcvalidity} we show that $\BPC((x,y),\cdot)$ is a maximal coupling of
$P(x,\cdot)$ and $P(y,\cdot)$.

\begin{prop}
	\label{prop:bpcvalidity}
	For any $\bar{Q} \in \Gamma(Q,Q)$, the output $(X,Y)$ of Algorithm \ref{alg:bpc} will follow a coupling ${\bar P \in \Gamma(P,P)}$. If $\bar{Q}$ is maximal then $\bar{P}$ will be as well.
\end{prop}

\vspace{-1em}
\begin{proof}
	At any point $x' \neq x$, $X$ will have density
	\eq{
		& q^m_{xy}(x') b_{xy}(x') + q^r_{xy}(x') c_{xy}(x') \\
		& = \big(q^m_{xy}(x') \wedge f(x,x') \big) + \big(q^r_{xy}(x') \wedge f^r_{xy}(x') \big) \\
		& = \big(q^m_{xy}(x') \wedge f(x,x') \big) + f^r_{xy}(x') = f(x,x').
	}
	The second equality holds because ${f^r_{xy}(x') \leq q^r_{xy}(x')}$.
	This in turn holds because $f(x, x') \leq q(x,x')$
	and ${q^m_{xy}(x') \leq q(x,x')}$ by Lemma~\ref{lem:qm}, so
	\eq{
		f^r_{xy}(x') & = {\big(f(x,x') \vee q^m_{xy}(x') \big) - q^m_{xy}(x')} \\
		& \leq q(x,x') - q^m_{xy}(x') = q^r_{xy}(x').
	}
	Integrating the density of $X$ over all $x' \neq x$ yields $\P(X = x) = {1 - \int f(x,x') \diff x'} = r(x)$, so we conclude $X \sim P(x,\cdot)$. A similar argument shows that ${Y \sim P(y, \cdot)}$. Thus $(X,Y)$ follows the desired coupling.
	
	If $\bar{Q}$ is maximal then by Lemma~\ref{lem:qm} the probability density at
	the proposal $(z,z) \sim \bar{Q}((x,y), \cdot)$ will be $q^m_{xy}(z)=q(x,z)\wedge q(y,z)$. 
	By the definition of $\bar{B}_1$, the probability of accepting a proposal $x' = y' =z$ for both $X$ and $Y$ will be
	\eq{
		\big( 1 \wedge \tfrac{f(x,z)}{q^m_{xy}(z)} \big) \wedge \big(1 \wedge \tfrac{f(y,z)}{q^m_{yx}(z)} \big)
		= \tfrac{f(x,z) \wedge f(y,z) }{q(x,z) \wedge q(y,z)}.
	}
	Combining this with the proposal density implies that the 
	overall coupled transition kernel density at ${X=Y=z}$ will be $f(x,z) \wedge f(y,z)$. Thus
	$\P(X=Y) = {\int f(x,z) \wedge f(y,z) \diff z}$, and by Lemma~\ref{lem:kernel_tv} we conclude that $\bar{P}$ is maximal.
\end{proof}

We observe that $\BPC$ matches the flexibility and computational efficiency of
$\BPSQ$ while offering a higher meeting probability at each iteration. The
ability to chose an arbitrary $\bar Q \in \Gamma(Q,Q)$ is also a significant
advantage of $\BPC$ over the full-kernel couplings. The extra effort required
to construct, validate, and draw from $\BPMR$ relative to $\BPMI$ shows how
challenging such refinements can be. Finally, we note that $\BPC$
can be more computationally efficient than the full-kernel couplings $\BPMI$ and
$\BPMR$, in that it avoids the `while' loops of Step 3 of Algorithms
\ref{alg:bpmi} and \ref{alg:bpmr}.

%%%%%

\section{Numerical Examples}
\label{sec:num}

%%%

\subsection{Biased Random Walk MH}
\label{sec:examples:expo}

\begin{figure}
	\centering
	\includegraphics[width=.95\linewidth]{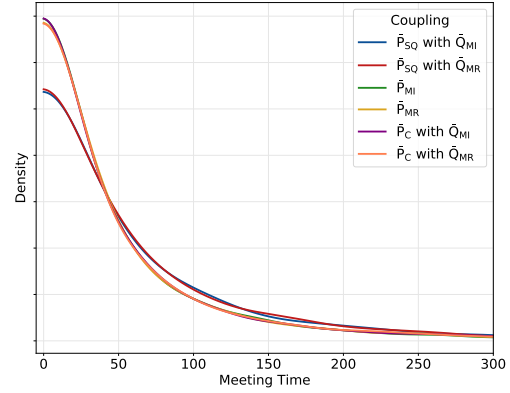}
	\caption{Distribution of meeting times for the example described in
		Section~\ref{sec:examples:expo}. The four maximal couplings yield almost
		identical distributions of shorter meeting times times while the two
		$\BPSQ$ methods yield almost identical distribution of longer ones. In
		this example the requirement that $\BPSQ$ accept meeting and non-meeting
		proposals at exactly the MH rate puts it at a significant disadvantage
		vs. the maximal couplings.
		\label{fig:meetingdist}}
\end{figure}

For our first example we consider a toy model that emphasizes the differences
between $\BPSQ$ and the maximal couplings $\BPMI$, $\BPMR$, and $\BPC$. We
assume an Exponential target distribution $\pi = \Expo(1)$ and draw proposals
from ${Q(z,\cdot) = \N(z + \kappa, \sigma^2)}$ with $\kappa > 0$. We then
accept or reject these proposals at the usual MH rate, so ${a(z, z') = 1 \wedge
	\exp\!\big( (z-z') (2 \kappa/\sigma^2 + 1) \big)}$. Our assumption on $\kappa$
implies $a(x,z) \wedge a(y,z) = a(x,z)$ if $x \leq y$ and $a(x,z) = 1$ if $z
\leq x$. Thus $Q$ will tend to propose increasing values while $a$ favors
decreasing ones. This tension between the proposal and target distributions
is characteristic of MH kernels that do not mix rapidly.

We construct the transition kernel couplings $\BPSQ$ with $\BQMI$, $\BPMI$,
$\BPMR$, and $\BPC$ with $\BQMI$ as described in Sections \ref{sec:cpl} and \ref{sec:cpl:fullkern}. We also
define a simple generalization of the maximal coupling with reflection residuals, 
such that either $x' = y'$ or ${y' - (y + \kappa) }= {(x + \kappa) - x'}$. 
We use the resulting $\BQMR$ as an alternative proposal kernel
coupling for $\BPSQ$ and $\BPC$. We set $\kappa = \sigma^2 = 3$, 
draw initial values independently from the target $\pi$,
and run 10,000 replications for each coupling option.
For each replication we record the meeting time ${\tau = \min(t \geq 0 : X_t = Y_t)}$.
As described in Section \ref{sec:intro}, such meeting times are of theoretical
and practical importance, and they also make a good measure of coupling performance.
We summarize the average behavior of these meeting times in Table~\ref{tab:times}
and present their full distribution in Figure~\ref{fig:meetingdist}.

\begin{table}[b]
	\caption{\normalsize Section~\ref{sec:examples:expo} Example Results} \label{tab:times}
	\begin{center}
		%\small
		\begin{tabular}{lcc}
			Coupling & Avg. Meeting Time & S.E. \\
			\hline \vspace{-8pt} \\ %space adjustment should be 8pt if we use \small
			$\bar P_{SQ}$ with $\bar Q_{MI}$ & 74.0 & 0.94 \\
			$\bar P_{SQ}$ with $\bar Q_{MR}$ & 75.6 & 0.99 \\
			$\bar P_{MI}$ & 60.5 & 0.84 \\
			$\bar P_{MR}$ & 60.9 & 0.87 \\
			$\bar P_{C}$ with $\bar Q_{MI}$ & 61.3 & 0.87 \\
			$\bar P_{C}$ with $\bar Q_{MR}$ & 62.2 & 0.89 \\
		\end{tabular}
	\end{center}
\end{table}

We find that both non-maximal couplings deliver average meeting times around 75
iterations, while the four maximal couplings deliver meeting times around 61
iterations. We recall that for a given state pair $(x,y)$, the maximal
couplings produce one value of $\P(X=Y \mid x,y)$ and two $\BPSQ$ couplings
produce another. The observed clustering of algorithms is consistent with the
idea that in this example meeting times are driven by these one-step meeting probabilities
rather than by behavior when meeting does not occur, 
which varies significantly by algorithm.

For a better understanding of these differences, we contrast the behavior of
$\BPSQ$ and $\BPC$.  Although we use the same underlying proposal coupling
${(x',y') \sim \bar Q((x,y),\cdot)}$ in each case, the two MH transition kernel
coupligns differ in that $\BPSQ$ accept its proposals at exactly the MH rate
while $\BPC$ uses a higher acceptance probability when $x'=y'$ and a lower one
when $x' \neq y'$. In this example, most proposals have a relatively low MH
acceptance probability to begin with. Thus $\BPC$ meets more quickly by
concentrating the little acceptance probability available on the draws where
they could result in a meeting $X=Y$, while $\BPSQ$ is forced to accept its
proposals at the same relatively low rate whether or not a meeting is proposed.

The above leads us to expect that maximal couplings might provide the greatest advantage over the status quo when low acceptance probabilities are typical, either due to the presence of a challenging target, an imperfect proposal distribution, or both. This example emphasizes simplicity over realism, but we would expect its principles to hold more broadly, especially in cases where mixing is relatively slow.

%%%

\subsection{Dimension Scaling with a Normal Target}
\label{sec:examples:mvn}

For our second example we consider MH on $\R^d$ with a target distribution ${\pi = \N(0, I_d)}$ and proposals $Q(z, \cdot) = \N(z, I_d\, \sigma^2_d)$. Following e.g. \citet{Roberts1997} and \citet{Christensen2005} we set $\sigma^2_d = \ell^2/d$ with $\ell = 2.38$. This example allows us to consider the role of the dimension and examine differences between the couplings when meeting probabilities are just one important aspect of their behavior.

\begin{figure}[t]
	\includegraphics[width=\linewidth]{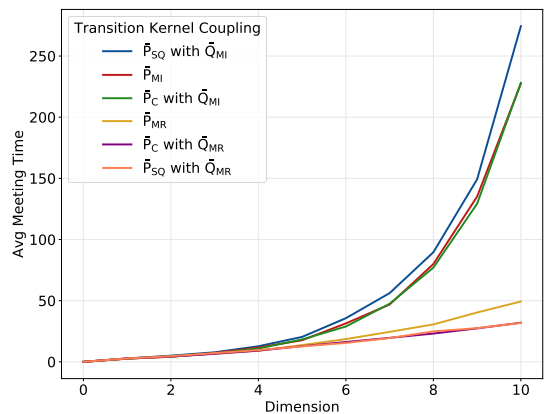}
	\caption{Average meeting times for a range of transition kernel couplings,
		as described in Section \ref{sec:examples:mvn}. Coupling strategies
		that involve reflections of the proposals appear
		to outperform the others; among the others, maximal couplings seem to perform
		better than non-maximal ones.
		\label{fig:meetingtimes}}
\end{figure}

As above, our main object of interest in this example is the number of iterations required for a pair of coupled chains to meet. We initialize chains on independent draws from $\pi$, consider dimensions $d = 1, \dots, 10$, and run 1,000 replications for each algorithm. We use a maximal coupling of acceptance indicators for $\BPSQ$ and $\BPC$, which appears to yield the best results among the simple acceptance indicator couplings.

We present the results of this experiment in Figure~\ref{fig:meetingtimes}. There we observe that $\BPSQ$ using $\BQMI$, 
$\BPC$ using $\BQMI$, and $\BPMI$ seem to yield meeting times that increase exponentially in dimension, although the maximal couplings outperform the non-maximal coupling. This blow-up is expected since these options involve independent or weakly dependent behavior when $X \neq Y$. The coupling $\BPMR$ delivers somewhat better behavior, with $\BPSQ$ and $\BPC$ with $\BQMR$ delivering the best performance. In higher dimensions it appears that the $\BQMR$ version of $\BPSQ$ may outperform its $\BPC$ counterpart, an interesting and perhaps counterintuitive result.

To understand these differences in meeting times, we must consider what happens under each coupling $\bar P$ when a meeting does not occur. For ${Q(z,\cdot) = \N(z, I_d \sigma^2_d)}$, any $\bar Q \in \Gamma^\mathrm{max}(Q,Q)$ will yield
${\P(x'=y' | x, y) = \P(\chi^2_1 \geq \lVert y-x\rVert^2/(4 \sigma^2_d))}$, see
e.g. \citet[chap.~3.3] {Pollard2005}. Since a meeting cannot occur unless
one is proposed, this expression is an upper bound on $\P(X=Y \mid x, y)$. Thus
we should expect the probability of $X=Y$ to fall off rapidly in $\lVert
y-x\rVert$, so that couplings which do not promote contraction between chains
in the absence of meeting will yield rapidly increasing meeting times.

\begin{figure}
	\includegraphics[width=\linewidth]{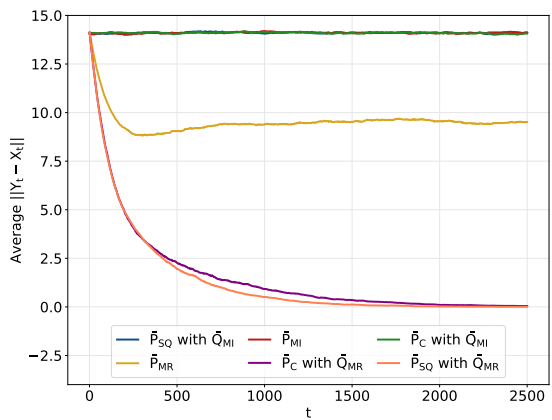}
	\caption{Distance between coupled MH chains by iteration, as described in
		Section \ref{sec:examples:mvn}. Couplings in which $X\neq Y$ implies
		$X,Y$ independent or almost independent display little contraction, while
		those based on the maximal reflection coupling of proposal distributions
		$\BQMR$ display strong contraction.
		\label{fig:avgdist}}
\end{figure}

Figure \ref{fig:avgdist} is consistent with these observations. As above we run 1,000 replications per algorithm, initializing each chain with an independent draw from the target distribution. We set $d=100$ to isolate the effects of contraction due to meeting from other contractive behavior and then track the average distance between chains as a function of iteration. 

We find that $\BPMI$, $\BPSQ$ with $\BQMI$, and $\BPC$ with $\BQMI$ produce little or no
contraction within the distance achieved by independent draws from the target. 
They then remain far enough apart that the probability of meeting is negligible. 
The coupling $\BPMR$ displays contraction up to a point. Finally we observe
that both the maximal and non-maximal couplings based on $\BQMR$ contract
rapidly to within a radius where meeting can occur. $\BPSQ$ appears to contract
more rapidly than $\BPC$, which is consistent with its slightly stronger performance in
high dimensions.

%%%%%

\section{Discussion}
\label{sec:discussion}

Couplings play a central role in the analysis of MCMC convergence and
increasingly appear in new methods and estimators.
Until now, no general-purpose algorithm has been available to sample from a maximal coupling
of MH transition kernels.
We fill this gap by introducing three such algorithms, which 
are implementable under the standard assumptions that one can draw 
proposals from a distribution $Q$ and compute the density $q$ 
and acceptance rate $a$ at any points of the state space. The proposed couplings are simple to apply and can be used with
a variety of MH strategies including the Metropolis-adjusted Langevin algorithm, 
pseudo-marginal methods, and the MH-within-Gibbs algorithm.

The experiments in Section \ref{sec:examples:expo} show that
the gains from using these methods can be large,
especially when there is a tension between the proposal density $q$ and the acceptance rate $a$.
On the other hand the example of Section \ref{sec:examples:mvn} shows
that maximality is sometimes less important than other properties of a coupling,
such as the contraction behavior when a meeting does not occur.
The two examples considered here are simple ones, and experiments with a wider range of MH algorithms
and target distributions would clarify the strengths and weaknesses of the proposed couplings.

This work raises several questions.  First, it it not known if
all couplings of MH kernels might be represented in the form of Algorithm~\ref{alg:bpc}
for some appropriate choice of proposal and acceptance couplings;
see \citet{nusken2019constructing} for the treatment of a
similar question in the setting of continuous-time Markov chains.
It would also be interesting to consider the use of sub-maximal proposal distribution couplings
in Algorithm~\ref{alg:bpc}, as suggested in the comment of \citet{Lee2020} on \citet{Jacob2020}.

Both meeting probabilities and contraction rates influence meeting times,
and one might wonder about deriving maximally contractive couplings
in analogy to the present work on meeting times.
Reflection couplings seem particular effective and 
are known to be optimal in special cases \citep{Lindvall1986}. 
In other cases, synchronous or `common random number' couplings yield
strong contraction \citep[e.g.][]{diaconis1999iterated}.  In most other scenarios the
user must construct a coupling tailored to the problem at hand.
The methods proposed here represent a step forward in coupling design, but many important questions remain.

From a more theoretical point of view, while the MH kernel $P$ is known to be the projection of the proposal $Q$ onto 
the set of $\pi$-reversible kernels in a certain metric \citep{billera2001geometric},
it is not known how $\bar{Q}$ relates to the set of maximal couplings
of $\pi$-reversible kernels. In particular, it would be interesting to know if the
strategy proposed in Section \ref{sec:cpl:conditional} corresponds to a projection of $\bar{Q}$
onto that set. 

Finally the couplings strategies mentioned above are all Markovian.
In some cases non-Markovian couplings are known to deliver more satisfactory
performance than Markovian ones
\citep[e.g.][]{smith2014gibbs}. The design of practical non-Markovian 
couplings for MCMC is a topic deserving further attention.

\subsubsection*{Acknowledgements}
The authors would like to thank Jun Yang, Persi Diaconis, and Niloy Biswas for helpful discussions.
Pierre E. Jacob gratefully acknowledges support by the National Science
Foundation through grants DMS-1712872 and DMS-1844695.

%%%%%%

\setlength{\itemindent}{-\leftmargin}
\makeatletter\renewcommand{\@biblabel}[1]{}\makeatother
\bibliography{maximal_mh_couplings}{}

%%%

\onecolumn

\appendix

\section{Appendix}
In Appendix~\ref{sec:app:bpmiproof}, we prove that
the coupling $\BPMI((x,y),\cdot)$ described in Algorithm \ref{alg:bpmi} is a maximal
coupling of $P(x,\cdot)$ and $P(y,\cdot)$ and analyze its computational
cost. Similarly, in Appendix~\ref{sec:app:bpmrproof} we prove that
the $\BPMR$ described in Algorithm~\ref{alg:bpmr} is valid and maximal
and analyze its cost.

\subsection{Validity, maximality, and computation cost of $\BPMI$}
\label{sec:app:bpmiproof}

We prove that Algorithm \ref{alg:bpmi} defines a coupling of the correct marginal distributions and that it attains the maximum one-step meeting probability.

\begin{prop}
	\label{prop:alg1validity}
	The draws $(X,Y)$ produced by Algorithm \ref{alg:bpmi} follow a maximal coupling of $P(x,\cdot)$ and $P(y,\cdot)$, with the property that $X$ and $Y$ are conditionally independent given $X \neq Y$. Moreover, the coupling probability is maximized among all possible couplings, so that $\BPMI((x,y), X= Y) = 1 - \lVert P(x,\cdot) - P(y, \cdot)\rVert_\TV.$
\end{prop}

\vspace{-1em}
\begin{proof}
	It suffices to show $Y$ has marginal distribution $P(y,\cdot)$ and the coupling probability equals ${1 - \lVert P(x,\cdot) - P(y, \cdot)\rVert_\TV}$.
	We define the residual measure of $y$ as $\tilde P_{yx}( \cdot ) \propto r(y) \delta_y(\cdot) + \tilde f_{yx}^r(\cdot)$, where $\delta_y$ is the point mass at state $y$ and $\tilde f_{yx}^r(y') \coloneqq f(y,y') -f(y,y')\wedge f(x,y')$ is the `unnormalized' $y$-residual density evaluated at $y'$.
	These definitions are consistent with the ones given in Section \ref{sec:cpl:bpml}. From this point of view, Step $3$ can be seen as a standard rejection sampler with proposal measure $P(y, \cdot)$ and target measure $\tilde P_{yx}( \cdot ) $. For any $y' \neq y$, let $f_{yx}(y')$ denote the transition density of the $Y$-chain from $y$  to $y'$. Then $f_{yx}(y')$  can be written as
	\begin{align*}
	f_{yx}(y') = f_{yx}(y', \text{Step } 2) + f_{yx}(y', \text{Step } 3).
	\end{align*}
	The first term works out to $f_{yx}(y', \text{Step } 2) = f(x,y')\wedge f(y,y')$,
	while the second term can be computed as 
	\begin{align*}
	& f_{yx}(y', \text{Step } 3) = \P(\text{Step 3})  f_{yx}(y'\mid \text{Step } 3)\\
	& =  (1 - \int f(x,z)\wedge f(y,z) dz) \frac 1 {c(x,y)} \tilde f_{yx}^r(y').
	\end{align*}
	Here $c(x,y) = 1 - \int f(y,z)\wedge f(x,z) dz$ is the normalizing constant of $\tilde P_{yx}( \cdot ) $,
	which also equals $r(y)+\int \tilde{f}^r_{yx}(z)dz$. 
	
	Putting all the terms together, we have $f_{yx}(y') = f(x,y')\wedge f(y,y') + \tilde f^r_{yx}(y')= f(y,y')$
	as desired, which justifies the validity of Algorithm \ref{alg:bpmi}.

    We can also observe that the coupling probability equals the probability
    that Algorithm \ref{alg:bpmi} stops at Step $2$. Therefore, the coupling
    probability satisfies 
    \begin{align*}
    \BPMI((x,y), X= Y) = \int f(x,y')\wedge f(y,y') dy'
    = 1 - \lVert P(x,\cdot) - P(y, \cdot)\rVert_\TV. 
    \end{align*}
    We conclude that Algorithm \ref{alg:bpmi} maximizes the
    coupling probability in one step.
\end{proof}

We analyze the computation cost of Algorithm \ref{alg:bpmi} as follows. To
draw one sample from Algorithm \ref{alg:bpmi}, one needs to run Step 1 once
with probability $1$,  Step 2 once with probability $1 - r(x)$,  Step 3 for $N$
times where $N$ is a random variable which equals $0$ with probability $1 -
\lVert P(x,\cdot) - P(y, \cdot)\rVert_\TV$, and otherwise  a Geometric 
random variable with success probability $ \lVert P(x,\cdot) - P(y,
\cdot)\rVert_\TV$. Meanwhile, Step $1$ contains one draw from $P$, Step $2$
contains two evaluations, Step $3$ contains one draw from $P$ and $0$ or $2$
evaluations, with probability $r(y)$ and $1-r(y)$ respectively.

Therefore, the expected number of draws from $P$ is $2$ and the expected number of
evaluations is $4 - 2r(x) - 2r(y)$. Taking account of the fact that each
draw from $P$ contains one draw from $q$ and one evaluation of the
acceptance ratio, then Algorithm \ref{alg:bpmi} contains $2$ draws from $q$ and
$6 - 2r(x) - 2r(y)$ evaluations in expectation. The variance of the computing
cost depends the total variation distance between $P(x,\cdot)$ and
$P(y,\cdot)$, and goes to infinity as $\lVert P(x,\cdot) - P(y,
\cdot)\rVert_\TV$ goes to zero.
This can motivate the consideration of sub-maximal coupling strategies,
as described in the comment of \citet{Lee2020} on \citet{Jacob2020}.

\subsection{Validity, maximality, and computation cost of $\BPMR$}
\label{sec:app:bpmrproof}

We  prove that Algorithm \ref{alg:bpmr} defines a valid coupling and attains the maximal coupling probability.

\begin{prop}
	\label{prop:alg2validity}
	The draws $(X,Y) \sim \bar{P}_\mathrm{MR}((x,y),\cdot)$ 
	produced by Algorithm \ref{alg:bpmr} follow a maximal coupling of $P(x,\cdot)$ and $P(y,\cdot)$.
\end{prop}

\vspace{-1em}
\begin{proof}
	As in the proof of Proposition \ref{prop:alg1validity},
	it suffices to show that $Y$ is distributed according to $P(y, \cdot)$ and that the meeting probability equals $1 - \lVert P(x,\cdot) - P(y, \cdot)\rVert_\TV$. Let the functions $f^m_{xy}$, $\tilde{f}^r_{xy} , \tilde{f}^r_{yx}$, and $\tilde{f}^t_{yx} $ have the definitions given in Section~\ref{sec:cpl:bpml}. For any $y' \neq y$,  the Markov transition density $f(y,y')$ can be written as the sum of three terms:
	\begin{equation}\label{eqn:bpmrtarget}
	f(y,y') = f^m_{xy}(y') +\tilde{f}^r_{yx}(y') \wedge \tilde{f}^r_{xy}(T_{yx}(y')) + \tilde{f}^t_{yx}(y').
	\end{equation}
	For any $y' \neq y$, let $f_{yx}(y')$ denote the transition density of the $Y$-chain from $y$  to $y'$ according to Algorithm \ref{alg:bpmr}. Then $f_{yx}(y')$ can also be written as the sum of three terms:
	\begin{equation}\label{eqn:bpmrtransition}
	f_{yx}(y') = f_{yx}(y', \text{Step } 2) + f_{yx}(y', \text{Step } 3(b)) + f_{yx}(y', \text{Step } 3(c)).
	\end{equation}
	We confirm that each term in Formula (\ref{eqn:bpmrtransition}) matches the corresponding term in (\ref{eqn:bpmrtarget}). For the first term, we have 
	\begin{align*}
	f_{yx}(y', \text{Step } 2) &= f(x,y') \P(Uf(x,y') \leq f(y,y'))
	= f(y,y')\wedge f(x,y').
	\end{align*} 
	For the second term, let $x' = T_{yx}(y')$ be the preimage of $y'$ through $T_{xy}$. It is not difficult to verify that $ T_{xy}^{-1} = T_{yx} $ and that the Jacobian of both $T_{xy}$ and $T_{yx}$  equals $1$. Thus the density of moving from $y$ to $y'$ through Step $3(b)$ will be
	\begin{align*}
	f_{yx}(y', \text{Step }3(b)) &= \tilde f_{xy}^r(x')  \P(U\tilde f_{xy}^r(x') \leq \tilde f_{yx}^r(y') ) \lvert J(T_{xy}^{-1})\rvert
	= \tilde{f}^r_{yx}(y') \wedge \tilde{f}^r_{xy}(T_{yx}(y')).
	\end{align*}
	This matches the second term in (\ref{eqn:bpmrtarget}). 

	Step $3(c)$ is again a rejection sampler with proposal $P(y,\cdot)$ and target ${ \tilde P^t_{yx}( \cdot ) \propto r(y) \delta_y(\cdot) + \tilde f^t_{yx}(\cdot)}$. Therefore, 
	\begin{align*}
	f_{yx}(y', \text{Step } 3(c))  & =  \P(\text{Step } 3(c)) \cdot f_{yx}(y'|\text{Step } 3(c))
	= \P(\text{Step } 3(c)) \frac {1}{\tilde c(x,y)} \tilde f_{yx}^t(y'),
	\end{align*}
    where $\tilde{c}(x,y) =1/ (r(y) + \int  \tilde f^t_{yx}(z) \diff z)$ is the normalizing constant of $\tilde P^t_{yx}(\cdot)$. 
	
	Meanwhile, we have 
	\begin{align*}
	\P(\text{Step } 3(c)) & = 1 - \int f^m_{xy} (y')dy'  -\int\tilde{f}^r_{yx}(y') \wedge \tilde{f}^r_{xy}(T_{yx}(y')) \diff y'\\
	& = r(y) + \int f(y,y') \diff y'-  \int f^m_{xy} (y') \diff y' -\int \tilde{f}^r_{yx}(y') \wedge \tilde{f}^r_{xy}(T_{yx}(y')) \diff y'\\
	& = r(y) + \int  \tilde f^t_{yx}(z) \diff z.
	\end{align*}
	Here the final equality uses Formula (\ref{eqn:bpmrtarget}). This yields 
	$ f_{yx}(y', \text{Step } 3(c))  = \tilde f_{yx}^t(y')$,
	which concludes the proof.
	
    We  also observe that the meeting probability equals the probability that
    Algorithm \ref{alg:bpmr} stops at Step $2$. Thus the coupling
    probability $\BPMR((x,y), X= Y) = \int f(x,y')\wedge f(y,y') dy' = 1 -
    \lVert P(x,\cdot) - P(y, \cdot)\rVert_\TV$ attains the upper bound given by the
    coupling inequality.
\end{proof}

The computation cost of Algorithm \ref{alg:bpmr} can be analyzed in a similar
way as the cost of Algorithm \ref{alg:bpmi}. We define
the one-step coupling probability $p_c := \int f_{xy}^m(y') \diff y' =
1- \lVert P(x,\cdot) - P(y, \cdot)\rVert_\TV $ and
the one-step rejection probability
$p_r := \tilde{f}^r_{yx}(y') \wedge \tilde{f}^r_{xy}(T_{yx}(y')) \diff y'$ .
To draw one sample from Algorithm
\ref{alg:bpmr}, one needs to run Step 1 once with probability $1$, run Step $2$
once with probability $1 - r(x)$, Step $3(a)$ once with probability $1 - p_c$,
and Step $3(b)$ once with probability  $1 - r(x) - p_c$. The number of runs of Step
$3(c)$ will be zero with probability $ 1- p_c - p_r$. Otherwise it will follow a
Geometric random variable with success probability $ 1 - p_c - p_r$.

 Meanwhile, Step $1$ contains one draw from $P$, Step $2$ contains two evaluations, Step $3(a)$ contains one evaluation, Step $3(b)$ contains two evaluations, Step $3(c)$ contains one draw from $P$ and $0$ or $2$ evaluations, with probability $r(y)$ and $1 - r(y)$ respectively. Therefore, the expected number of draws from $P$ is $2$, and the expected number of evaluations is $7 - 4r(x) - 2r(y) - 3p_c$. If one takes into account the fact that each draw from $P$ itself contains one draw from $q$ and one evaluation of the acceptance ratio, then Algorithm \ref{alg:bpmi} contains $2$ draws from $q$ and $7 - 4r(x) - 2r(y) - 3p_c$ evaluations, in expectation. This is greater than the expected cost of Algorithm~\ref{alg:bpmi} by $3 - 3p_c - 2r(x)$. (This quantity is  between $0$ and $3$ as $p_c + r(x)  \leq 1$.) The variance of the computation cost also depends the total variation distance between $P(x,\cdot)$ and $P(y,\cdot)$, and goes to infinity as $\lVert P(x,\cdot) - P(y, \cdot)\rVert_\TV$ goes to zero, as noted above.

\end{document}